# PROPERTIES OF GRAPHENE-RELATED MATERIALS CONTROLLING THERMAL CONDUCTIVITY OF THEIR POLYMER NANOCOMPOSITES

Samuele Colonna, Daniele Battegazzore, Matteo Eleuteri, Rossella Arrigo, Alberto Fina[*]

Dipartimento di Scienza Applicata e Tecnologia, Politecnico di Torino, Alessandria Campus, Viale Teresa Michel 5, 15121 Alessandria, Italy

## Abstract

Different types of graphene-related materials (GRM) are industrially available and have been exploited for thermal conductivity enhancement in polymers. These include materials with very different features, in terms of thickness, lateral size and composition, especially concerning the oxygen to carbon ratio and the possible presence of surface functionalization. Due to the variability of GRM properties, the differences in polymer nanocomposites preparation methods and the microstructures obtained, a large scatter of thermal conductivity performance is found in literature. However, detailed correlations between GRM-based nanocomposites features, including nanoplatelets thickness and size, defectiveness, composition and dispersion, with their thermal conductivity remain mostly undefined.

In the present paper, the thermal conductivity of GRM-based polymer nanocomposites, prepared by melt polymerization of cyclic polybutylene terephtalate oligomers, exploiting 13 different GRM grades, was investigated. The selected GRM, covering a wide range of specific surface area, size and defectiveness, secure a sound basis for the understanding of the effect of GRM properties on the thermal conductivity of their relevant polymer nanocomposites. Indeed, the thermal conductivity obtained appeared to depend on the interplay between the above GRM feature. In particular, the combination of low GRM defectiveness and high filler percolation density was found to maximize the nanocomposites thermal conductivity.

## Keywords

---

[*] Corresponding author. E-mail: alberto.fina@polito.it (Alberto Fina)

## Abbreviations

Cyclic butylene terephthalate (CBT), graphene-related materials (GRM), graphite nanoplatelets (GNP), graphene oxide (GO), reduced graphene oxide (RGO), few-layer graphene (FLG), multi-layer graphene (MLG), Thermally Reduced GO (TRGO), Field Emission Scanning Electron Microscope (FESEM)

# 1. Introduction

The management of heat transport has progressively become a challenging problem owing to the miniaturization of products, especially in electronics [1]. On the other hand, the need to enhance energy efficiency in all industrial, transport and buildings applications is driving research and development on thermally efficient materials and devices. With this background, the remarkable corrosion resistance, the ease of processing, lightweight and low-cost of polymer matrices attracted attention for their exploitation in heat exchangers in order to replace metallic parts [2-4]. Furthermore, the use of polymer-based heat exchanger in principle opens the possibility for the manufacturing of flexible heat exchangers, that are particularly relevant to applications in flexible electronics [1] as well as in wearable and implantable devices [5, 6]. Albeit polymers are well known for their poor heat conduction properties (with typical values ranging between 0.1 and 0.5 W m$^{-1}$ K$^{-1}$ [2, 4]), the inclusion of particles or nanoparticles with intrinsically high thermal conductivity may lead to the formation of thermally conductive composites or nanocomposites [4, 7, 8]. Different types of particles/nanoparticles, including graphite, carbon nanotubes, diamond, metal micro- and nanoparticles, boron nitride and their combinations can be exploited for the enhancement of polymer thermal conductivity. In the last decade, significant research attention was focused on graphene and related materials. Indeed, different studies reported thermal conductivity values for pristine graphene between 2000 and 5000 W m$^{-1}$ K$^{-1}$, depending on the measurement technique [9-14]. However, this outstanding heat conductivity is affected by several parameters, including defectiveness [12, 15, 16], lateral size [17], number of layers [9, 18] and presence of wrinkles [11, 19].

Despite the development of several synthesis techniques [20, 21], the production of high quality monolayer graphene is still limited at small scale, mainly by mechanical cleavage [22], chemical vapour deposition [23] and epitaxial growth on SiC [24]. On the other hand, larger-scale production of graphene and related materials is possible by other techniques, including oxidation or salification of graphite followed by chemical [25, 26] or thermal reduction [27, 28], electrochemical exfoliation of graphite [29-31] and liquid phase exfoliation of graphite [32-34]. All of these techniques can lead to the production of monolayer graphene, but with limited yield. Besides, these allow for a large scale production of the so-called graphene-related materials (GRM), including graphite nanoplatelets (GNP), graphene oxide (GO), reduced graphene oxide (RGO), few-layer graphene (FLG) and multi-layer graphene (MLG) [35]. Thermal conductivity of GRM may be significantly lower than for graphene, owing to the higher number of layers and/or the presence of oxidized carbons or other defects, acting as scattering points for phonon transmission [9, 14, 36]. However, GRM have been widely exploited for the enhancement of polymers thermal conductivity [37], while sometimes mistermed to as graphene.

The implementation of the heat conduction property in polymer nanocomposites requires a proper control of the nanoparticles organization within the organic matrix and effective interfaces, to guarantee for thermal transfer between conductive nanoparticles as well as between nanoparticles and the surrounding matrix. The functionalization of GRM, either covalent or non-covalent, may help in both the enhancement of nanoplatelets/polymer interaction and the nanoparticles dispersion [38]. However, the covalent functionalization of graphene and GRM modifies the conjugation of their structure, increasing materials defectiveness, introducing $sp^3$ carbons that act as phonon scattering points [39] and consequently decreasing the intrinsic thermal conductivity of the nanoplatelets [40]. On the other hand, chemical functionalization on graphene and related materials have been proposed and exploited for the enhancement of thermal transfer at interfaces between GRM as well as between GRM and the polymer matrix in nanocomposites. The concept of molecular junctions between GRM was widely studied by different computational tools [41-43] and exploited in a few experimental papers [44, 45] to reduce thermal resistance at contacts between conductive GRM flakes. In polymer nanocomposites, the strategy of enhancing thermal interface between the conductive particles and the surrounding matrix was also widely exploited. Computational studies by molecular dynamics [46-48] indeed suggested an increase in thermal conductance when short chains covalently linked to the GRM are strongly interacting with the polymer matrix. Both covalent and non-covalent functionalization were in fact exploited to promote dispersion of GRM as well as possibly enhancing the thermal conductance at interfaces, demonstrating significantly enhanced thermal conductivity for nanocomposites prepared with functionalized GRM, compared with their pristine counterparts [49-52]. Unfortunately, the relatively complex functionalization of GRM is currently a limiting factor for the large-scale application. Beside the interfacial interaction between GRM and polymers, the organization of the nanoparticles can also be tailored through the processing route [37, 53]. Indeed, several methods can be exploited for the preparation of polymer/GRM nanocomposites, including solvent mixing, melt-mixing, in-situ polymerization and latex mixing, each with its own pros and cons in terms of dispersion, cost effectiveness, processing times, production scale [38]. Melt blending process are clearly advantageous in terms of cost and industrial scalability, but dispersion of non-functionalized GRM by direct compounding in molten polymer remains challenging [54, 55], both owing to the limited chemical affinity and the high viscosity. Polymerization during extrusion was recently demonstrated to be a promising route for efficient dispersion of GRM, despite limited to polymers obtainable by ring opening polymerization, such as polybutyleneterephthalate [56-60], polylactic acid [61, 62] and polyamide 6 [63, 64].

The above-summarized scenario explains the wide scatter of thermal conductivity enhancement reported in literature for polymer based GRM nanocomposites, which relates to the large differences in GRM structure and properties, as well as the type of polymer, the nanocomposite preparation method and the nanostructure obtained. While a comprehensive review of literature for thermal conductivity of polymer nanocomposites embedding GRM is beyond the scope of this paper, an overview of thermal conductivity enhancement factor ($\lambda/\lambda_{matrix}$) for a number of GRM-based nanocomposites [50, 51, 57, 58, 65-102] is reported in Figure 1, evidencing a severe scatter of results.

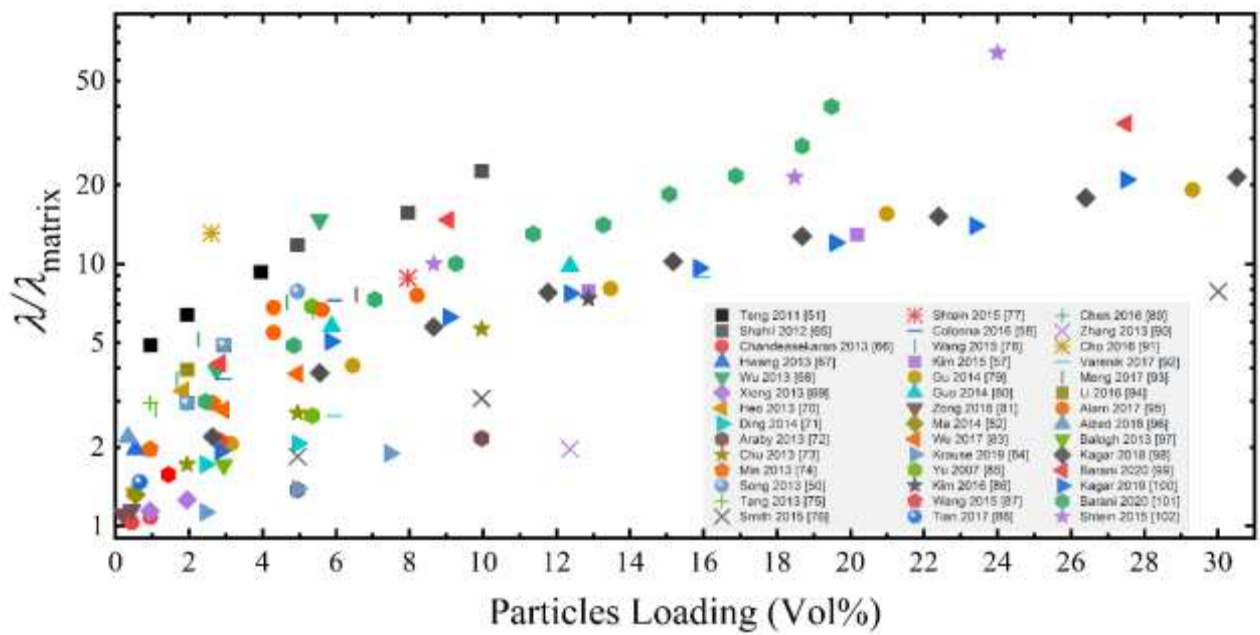

*Figure 1: Literature review for thermally conductive polymer nanocomposites embedding graphene-related materials. Data are taken from references [50, 51, 57, 58, 65-102].*

In this frame, detailed correlations between the structural features of GRM (e.g. lateral size, thickness, oxidation, defectiveness) and the thermal conductivity of corresponding polymer nanocomposites remain mostly undefined. Indeed, only a few papers reported comparisons between nanocomposites prepared in the same conditions using different grades of GRM. Yu *et al.* demonstrated the thermal conductivity of epoxy-based GNP nanocomposites to increase as the nanoplates size was reduced and the aspect ratio increased, at a constant nanoparticles loading [85]. Shtein *et al.* also reported epoxy-based nanocomposites with a series of GNP grades differing for lateral size, thickness and defect density, showing best thermal conductivity to be obtained with large lateral dimensions and low defectiveness [77]. Kim *et al*. prepared polycarbonate-based nanocomposites with different grades of GNP to correlate the nanocomposites thermal conductivity to the lateral size and thickness of GNP [86]. In agreement with previous papers, GNP with larger lateral size and thickness were found to deliver higher thermal conductivity in composites, whereas a minor effect of differences in GNP defectiveness (assessed by Raman scattering) was reported.

The previously reported papers compared thermal properties obtained with GNP grades fully comparable in terms of surface chemistry, selected to minimize differences in the particle/matrix interaction and dispersability. However, a more comprehensive comparison between different types of graphene related materials, obtained from different methods and therefore having different surface chemistry and structural properties, may provide additional information on the correlation between the different GRM structural features. With this aim, 13 grades of graphene-related materials, from different producers, were selected in this work, representing a wide span in lateral size, surface area, defectiveness and production technology. These were processed by melt extrusion polymerization of cyclic butylene terephthalate oligomers, producing nanocomposites with variable performances, which allowed identifying a correlation between nanocomposites thermal conductivity, the GRM structural features and their organization into a percolating network.

## 2. Materials and Methods

### 2.1. Materials

Cyclic butylene terephthalate oligomers [CBT100, Mw = $(220)_n$ g/mol, $n$ = 2-7, melting point= 130 ÷ 160 °C] and butyltin chloride dihydroxide catalyst (96%, melting point 150 °C,) were purchased from IQ-Holding[†] (D) and Sigma-Aldrich (I), respectively. Acetone (99+%) was purchased from Alfa Aesar (I).

Different types of GRM were selected for the present study, among materials with relatively large commercial availability and focusing on low surface oxidation. Selection was neither aimed to be comprehensive nor to represent all the different production methods and producers, but was exclusively carried out to cover a wide span GRM structural features, including lateral size, surface area and defectiveness. The main structural features of the selected GRM are summarized in Table 1.

---

[†] Distributor of products previously commercialized by Cyclics Europe GmbH

It is worth mentioning that, independently on the terminology used by the producers of the commercial products addressed here, all the different grades are considered to be graphite nanoplates, accordingly with terminology proposed by Bianco *et al*. [35] and will be referred in this work to as GRM-#, as assigned in Table 1. All the materials were used as received. Furthermore, GRM-4 and GRM-5 were selected to undergo annealing for 1 h at 1700 °C at ~ 50 Pa in a vacuum oven (Pro.Ba., I) to decrease the structural defectiveness, as previously reported [58]. High-temperature annealed material are referred to as GRM-4T and GRM-5T.

Table 1. Code used in the article, method of production, grade and producer, BET, Raman $I_D/I_G$ ratio, Oxygen content calculated from XPS of the different GRM.

| GRM-# | Preparation method | Grade & Producer | Surface Area BET [$m^2/g$] | Defectiveness, as $I_D/I_G$ ratio by Raman | oxygen %, atomic by XPS |
|---|---|---|---|---|---|
| 1 | EXP | Research grade [58], Avanzare (E) | 22 | 0.12 | 1.8 |
| 2 | EXP | Research grade [59], Avanzare (E) | 39 | 0.10 | 5.0 |
| 3 | ND | G2 Nan, Nanesa (I) | 27 | 0.08 | 1.7 |
| 4 | TRGO | Research grade [15], Avanzare (E) | 210 | 1.58 | 3.2 |
| 4T | TRGO + HT | GRM-4, annealed 1700°C | 123 | 0.09 | 0.4 |
| 5 | TRGO | EXG98, Graphit Kropfmühl (D) | 450 | 0.92 | 7.0 |
| 5T | TRGO + HT | GRM-5, annealed 1700°C | 300 | 0.10 | 1.1 |
| 6 | TRGO | Research grade [103], Avanzare (E) | 196 | 1.35 | 2.0 |
| 7 | LPE | ElicarbSP8082, Thomas Swan (UK) | 35* | 0.22 | 4.4 |
| 8 | ND | x-GnP-M25, XG Science (US) | 98 | 0.16 | 0.5 |
| 9 | ND | A12, Graphene Supermarket (US) | 38 | 0.40 | 4.9 |
| 10 | ND | x-GnP-C500, XG Science (US) | 500* | 0.68 | 5.7 |
| 11 | ND | N002-PDR-HD, Angstron Materials (US) | 406 | 1.33 | 1.0 |

\* As from product technical datasheet

EXP: expansion of graphite intercalation compound

TRGO: oxidation of graphite followed by thermal reduction

LPE: Liquid phase exfoliation of graphite

HT: High Temperature annealing

ND: non disclosed

## 2.2. Nanocomposites preparation

The nanocomposites preparation followed a previously developed method [58, 59] at 5 wt.% nanoplates loading. Briefly, it consisted in a preliminary dissolution of CBT in acetone (~ 120 ml) for 2 hours under vigorous stirring followed by the addition of the required amount of GRM and manual mixing. Then, the mixture was firstly stored in a chemical hood (~ 2h), to allow solvent evaporation, and subsequently dried at 80°C for 8 hours under vacuum (~$10^1$ mbar), to extract residual acetone and moisture. In a second step, the dried mixture was pulverized, loaded into a co-rotating twin screw micro-compounder (DSM Xplore 15, NL) and melt mixed for 5 min at 100 rpm and 250 °C. Then, 0.5 wt.% of butyltin chloride dihydroxide (with respect to the oligomer amount) was added and the extrusion carried on for further 10 minutes at 100 rpm to complete CBT polymerization into pCBT. A nitrogen flux was used during the whole process to restrain thermoxidative degradation and hydrolysis of polymer.

## 2.3. Characterization

Raman spectroscopy was carried out on a inVia Reflex (Renishaw PLC, UK) microRaman spectrophotometer equipped with a cooled charge-coupled device camera, directly on powder deposited on a glass slide. Samples were excited with a diode laser source (514.5 nm), with a power of 10 mW. The spectral resolution and integration time were 3 cm$^{-1}$ and 10 s, respectively. For each of the selected GRM grades, spectra were acquired in multiple points. The $I_D/I_G$ ratio were calculated for each spectrum (after careful baseline subtraction) and averaged values are reported with their standard deviation. For each of the GRM grades, a representative spectrum was obtained by averaging multiple spectra and finally normalizing on the most intense peak.

Morphological characterization of both nanoparticles and nanocomposites was performed by a high resolution Field Emission Scanning Electron Microscope (FESEM, MERLIN 4248 by ZEISS, D) operated at 5kV. GRM, deposited on a conductive tape, were directly observed without any further preparation. Nanocomposites were cryofractured in liquid nitrogen, then coated with about 5 nm of chromium sputtered layer to avoid electrostatic charge accumulation on the surface.

Rheological properties of pCBT/GNP nanocomposites were measured on a strain-controlled rheometer (Advanced Rheometric Expansion System, ARES, TA Instruments, USA) equipped with a convection oven to adjust the temperature. Tests were carried-out in parallel-plate geometry on disks with thickness and diameter of 1 and 25 mm, respectively. Before measurements, specimens were dried at 80 °C in vacuum for 8 h. Oscillatory frequency sweeps ranging from 0.1 to 100 rad/s with a fixed strain (chosen and selected for each sample in order to fall in the linear viscoelastic region) were performed in air at 240 °C, to investigate the viscosity and yield stress of the nanocomposites. After the sample loading, an approximate 5 min equilibrium time was applied prior to each frequency sweep. A Carreau-like model, modified with a term accounting for yield stress behavior [97] was exploited to differentiate the nanocomposites microstructure:

$$|\eta^*(\omega)| = \frac{\sigma_0}{\omega} + \frac{\eta_0^*}{[1+\tau\omega]^{1-n}}$$

where $\eta_0^*$ is the zero-shear viscosity, $\tau$ is the characteristic relaxation time of the polymer, $n$ is the flow index and $\sigma_0$ is the yield stress. In this equation, the first term describes the yield stress (i.e. the maximum shear stress that the polymer can experience before starting to flow) behavior at low frequencies, while the second term represents the classical trend of the viscosity curve predicted by Carreau model.

Isotropic thermal conductivity was evaluated by transient plane heat source method (ISO 22007-2, using a TPS 2500S by Hot Disk AB, S) equipped with a Kapton sensor (radius 3.189 mm). Specimens were hot-pressed (250 °C) into disks with 4 mm thickness and 15 mm diameter. Before testing, these were stored in a constant climate chamber (Binder KBF 240, D) at 23.0 ± 0.1 °C and 50.0 ± 0.1 %R.H. for at least 48 h. The measuring setup was inserted in a container dipped into a silicon oil bath (Haake A40, Thermo Scientific Inc., USA) equipped with a temperature controller (Haake AC200, Thermo Scientific Inc., USA) in order to control the temperature at 23.00 ± 0.01 °C.

## 3. Results and Discussion

### 3.1. Effect of the GRM defectiveness

The defectiveness of GRM is known to decrease their thermal transport properties and is expected to affect the final heat conduction properties of their relevant polymer nanocomposites. Different methods have been used to evaluate the defectiveness of GRM, including, thermogravimetric analysis [104], X-ray diffraction [15, 105], X-ray photoelectron spectroscopy [104, 106] and, most importantly, Raman spectroscopy [107, 108], the latter being the only technique able to provide a wealth of information related to GRM quality in a few minutes. Graphene and its related materials are typically characterized by three main peaks: the G band at ~ 1580 cm$^{-1}$, the D band at ~ 1350 cm$^{-1}$ and the G' band at ~ 2700 cm$^{-1}$ [107, 109]. In particular, the $I_D/I_G$ ratio is the simplest and most used defectiveness indicator: indeed, typically, the higher the ratio, the higher the amount of the defects [16]. However, it is worth noting that when the distance ($L_D$) between two consecutive defects is lower than ~ 3 nm a reduction in the $I_D/I_G$ ratio was observed experimentally and calculated theoretically, despite the increasing amount of defects [107]. Another defectiveness indicator is the ratio between the intensity of the D band and the D' band (a peak appearing at ~ 1621 cm$^{-1}$ in defective materials) proposed by Eckmann *et al.* [110] as an indicator of the defect type, *i.e.* boundary, vacancy or sp$^3$ carbon. Finally, the shape of the G' peak might help evaluating the number of layers in multilayer graphene, but applicability is limited to a few layers [109] and therefore this method cannot be exploited in assessing thickness of GNP. Raman spectra for the different GRM exploited in the present work and the $I_D/I_G$ ratios are shown in Figure 2. For all the materials the G' peak is the convolution of two main peaks, centred at ~ 2690 cm$^{-1}$ and ~ 2725 cm$^{-1}$, that are typical for graphene-related materials with more than 5 graphene layers [108]. Most of the selected GRM show intense G peak and low-intensity D band ($I_D/I_G$ ratio < 0.2), confirming these as relatively low defective materials, accordingly with the aim of our GRM selection. However, GRM-4, -5, -6, obtained by oxidation of graphite followed by thermal reduction (TRGO) showed an expectedly higher $I_D/I_G$ ratio, as well as GRM-10 and GRM-11, with $I_D/I_G$ ratio close to or higher than 1. The intense D band and the appearance of D' as a shoulder at ~ 1620 cm$^{-1}$, coupled with the broader G' peak, evidence for the presence of defects, typically remaining after incomplete thermal reduction and corresponding primarily to sp$^3$ carbons and vacancies. On the other hand, after thermal annealing at 1700 °C, dramatic changes in the Raman spectra are observable, with a sharpening of both the G and G' peaks as well as the reduction of the intensity of the D-peak, as previously reported [15, 58].

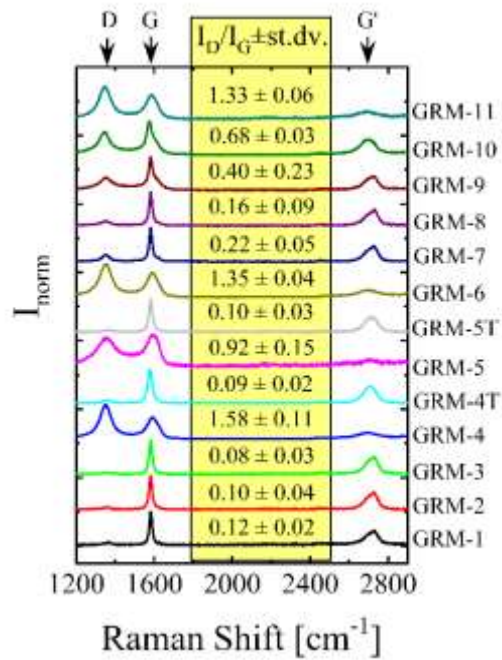

*Figure 2. Representative Raman spectra for the selected GRM, normalized with respect to the highest intensity peak. $I_D/I_G$ ratio and standard deviation are also reported for each GRM.*

The relationship between the defectiveness of the graphene-related materials exploited in this work and their nanocomposites thermal conductivity (at 5 wt.% GRM content, Table 2) is shown in Figure 3. At the selected GRM concentration, enhanced heat conduction properties, up to 4 times the value of the neat polymer (0.24 W m$^{-1}$ K$^{-1}$), were observed for most of the prepared materials, with best value close to 1.0 W m$^{-1}$ K$^{-1}$ for pCBT + 5% GRM-5T. Grouping the GRM in different families demonstrate that best materials for the improvement of the thermal conduction properties of pCBT are high temperature annealed TRGO. Indeed, the GRM-4T and GRM-5T exhibit $\lambda$=0.89 and 0.99 W m$^{-1}$ K$^{-1}$, respectively, with $I_D/I_G$ ratio $\approx$ 0.1, outperforming all formulations based on the other GRM grades. Based on these results, one might conclude that the lower defectiveness may correspond to the higher heat conduction property of composites. However, it is worth observing that the nanocomposite GRM-8, having one of the lowest $I_D/I_G$ ratio (~ 0.16), exhibited a rather low thermal conductivity value of 0.46 W m$^{-1}$ K$^{-1}$. Therefore, the effect of the GRM defectiveness is clearly not sufficient to explain thermal conductivity of the relevant nanocomposites. Indeed, while nanocomposites containing nanoplates with similar morphology and aspect ratio exhibit higher thermal conductivity at lower $I_D/I_G$ ratio, as previously reported [58, 60], the exploitation of different GRM with similarly low defectiveness is necessary but not sufficient to achieve high thermal conductivity.

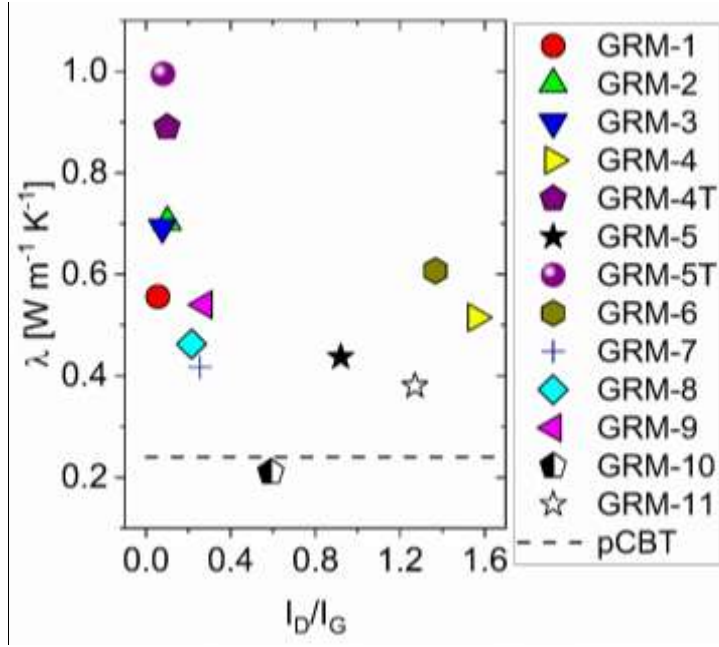

*Figure 3. Comparison between the nanocomposites thermal conductivity and GRM defectiveness (evaluated as $I_D/I_G$ ratio)*

*Table 2: Thermal conductivity and rheological parameters of pCBT GRM nanocomposites*

| Sample name | | $\lambda$ [W m$^{-1}$K$^{-1}$] | $\sigma_0$ | $R^2$ |
|---|---|---|---|---|
| pCBT | | 0.24 ± 0.01 | 0.3 | 0.999 |
| pCBT 5% GRM-# | 1 | 0.56 ± 0.01 | 0.1 | 0.985 |
| | 2 | 0.70 ± 0.01 | 0.8 | 0.995 |
| | 3 | 0.69 ± 0.01 | 1.2 | 0.991 |
| | 4 | 0.51 ± 0.01 | 8.8 | 0.997 |
| | 4T | 0.89 ± 0.01 | 53.9 | 0.992 |
| | 5 | 0.44 ± 0.01 | 43.6 | 0.999 |
| | 5T | 0.99 ± 0.01 | 72.8 | 0.998 |
| | 6 | 0.61 ± 0.01 | 24.8 | 0.995 |
| | 7 | 0.42 ± 0.01 | 0.6 | 0.999 |
| | 8 | 0.46 ± 0.01 | 2.7 | 0.857 |
| | 9 | 0.54 ± 0.01 | 8.4 | 0.996 |

| | 10 | 0.21 ± 0.01 | 2.7 | 0.998 |
| | 11 | 0.38 ± 0.01 | 9.1 | 0.898 |

## 3.2. Effect of the GRM particle size and morphology

The aspect ratio of nanoparticles was previously reported to affect the thermal conductivity of both graphene [9] and GRM-based polymer nanocomposites [77, 85, 86], as well as polymer/graphene laminates [111, 112]. However, a reliable and representative quantification of average values and distribution of lateral size and thickness to determine the aspect ratio of a GRM powder remains challenging. In fact, while electron microscopies and scanning probe microscopies have been widely used to measure the size of GRM flakes [20], the procedure applied in the preparation of the sample to be observed may result in some size selection of the flakes. For instance, suspending GRM in solvent via high shear mixing or sonication, followed by drop casting deposition, may result in a sample representing mainly the finer fraction, especially when starting from particles with a very broad size distribution. Even more importantly, the average size and distribution of the particles in as-received GRM powders may not be representative of the particle size once embedded in a polymer nanocomposite. Indeed, further fragmentation and/or exfoliation may easily occur, especially during energy-intensive procedures applying sonication steps and/or high shear, such as in melt compounding of polymer/GNP nanocomposites [60, 113]. For these reasons, the lateral size and thickness of the pristine GRM used were not quantified in this work. However, FESEM was routinely carried out to get a qualitative characterization of the as received nanoparticles and highlight the significant structural difference among the selected GRM grades. Representative micrographs showing morphologies for the different graphene-related materials are reported in Figure 4.

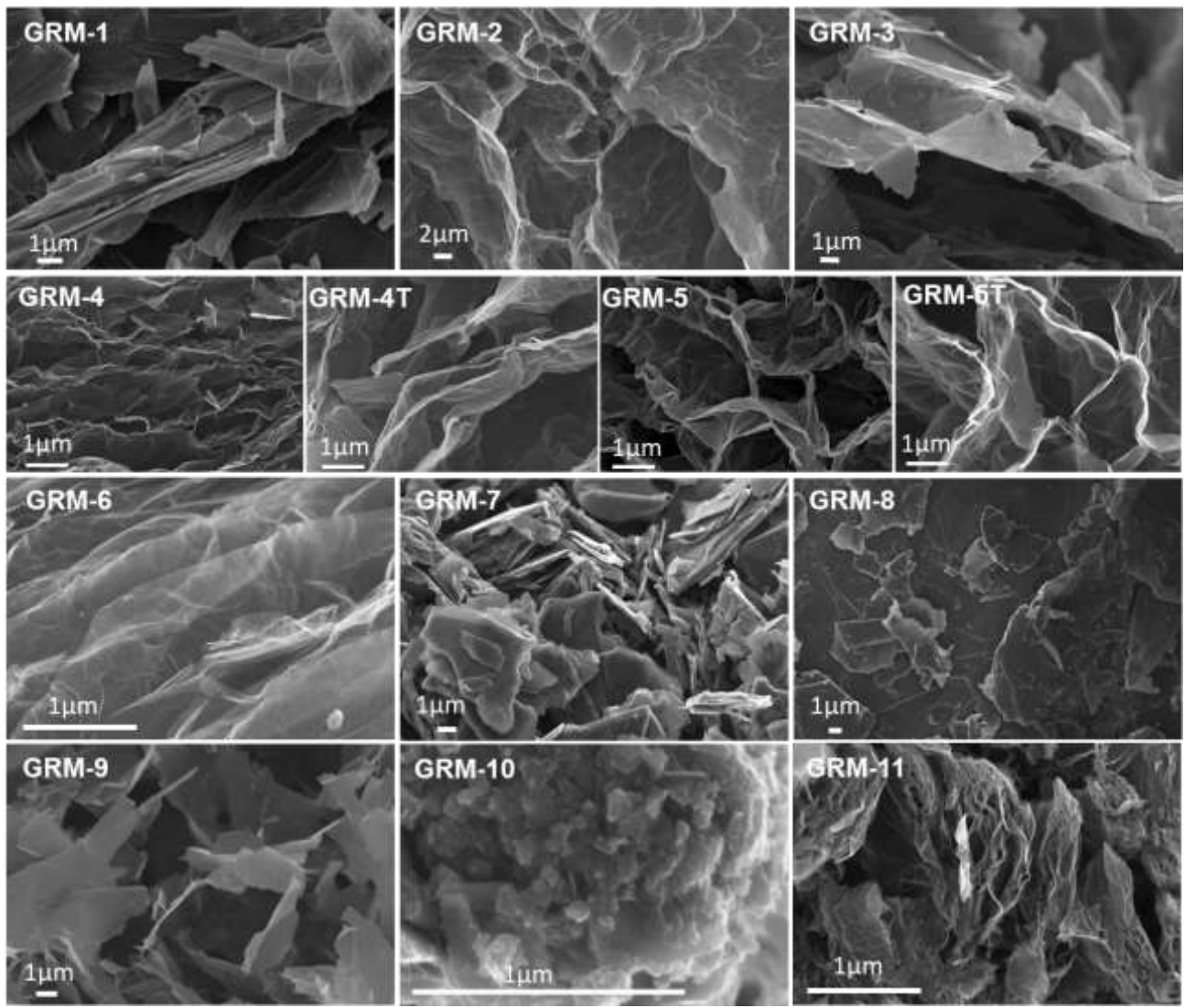

*Figure 4. Representative FESEM micrographs for the different GRM exploited in the present work. Different magnifications were selected to best highlight the distinctive features of the different grades.*

GRM-1 is characterized by flat and highly stacked nanoplates, in the range of ≥ 10 nm thickness and typically 5 to 10 µm lateral size. GRM-2 showed a completely different structure, characterized by well-separated large (tens of µm) and wavy thin layers of a few nm thickness, clearly obtained during the expansion of the graphite intercalation compound. GRM-3 is constituted by flat and apparently rigid flakes in the range of several microns microns (typical 5 to 20 µm) and thickness in the range of 10 nm, partially stacked in larger aggregates. GRM-4 is characterized by a well-expanded accordion-like structure, with a few nm thick layers, which appears to be further expanded after the high temperature annealing (GRM-4T). Similar structures were observed for GRM-5/5T, with some higher expanded and slightly more crumpled structure, as compared to GRM-4/4T. GRM-6 also appears as large (tens of µm) and very thin layers, as suggested by the marked transparency in FESEM observation. GRM-7 is characterized by very flat and apparently rigid flakes, with lateral size mostly in the range 1 to 5 µm and thickness up to 100 nm. GRM-8 is constituted by nanoparticles with a large distribution of both lateral size (2 to 30 µm) and typical thickness of a few tens of nanometres. GRM-9 appears similar to GRM-3, with lateral size typically in the range of 3 to 10 µm and approx. 10 nm thick. GRM-10 is constituted by submicronic flakes, typically in the range of a few hundreds of nm, very strongly aggregated in large agglomerates. Finally, GRM-11 is characterized by a clear accordion-like structure, apparently similar to GRM-4.

The thermal conductivity values for the pCBT nanocomposites containing 5 wt.% of each graphene-related materials (Table 2) show a general enhancement heat conduction respect to the neat polymer matrix, except for GRM-10. Indeed, in the latter case, the conductivity is even slightly lower than for pristine pCBT, suggesting the tiny GRM flakes to be unable to build a thermally effective network. Nanocomposites with larger lateral size and flat particles (GRM-1, -3, -7, -8, -9) exhibited thermal conductivity values ranging between 0.42 and 0.69 W m$^{-1}$ K$^{-1}$, higher performance corresponding to higher apparent aspect ratio in the as received GRM. The exploitation of highly expanded GRM (# 2, 6, 4/4T, 5/5T and 11), characterized by wavy thin flakes, also resulted in a broad thermal conductivity range (0.44 to 0.99 W m$^{-1}$ K$^{-1}$) for their nanocomposites. This wide span in $\lambda$ appears to be mainly related to the degree of defectiveness rather than to the limited morphological differences that can be appreciated by electron microscopy. Despite previous literature reported higher nanocomposite thermal conductivity when exploiting flat GRM instead of wavy GRM [57, 73], the same trend cannot be confirmed here and the waviness of pristine particles does not appear to be relevant as such. However, the presence of wrinkles, folds and twists may correlate with both the dispersability in polymers and the fragmentation during mixing.

Overall, limited correlations may be drawn between the as received particle size and/or morphology and the thermal conductivity of the nanocomposites, owing to the complexity of de-aggregation and fragmentation phenomena occurring during nanocomposites preparation. The combination of the processing method and the morphological features of as received GRM is indeed determining the organization of nanoparticles within the composite, which is discussed in the following section 3.3.

### 3.3. Effect of nanoflakes organization in the polymer matrix

Several techniques may provide information on the dispersion, distribution and organization of nanoparticle network within the polymer matrix. The most commonly used techniques include scanning and transmission electron microscopy, micro X-ray computed tomography (micro-CT) and rheology, each of these techniques having its own strengths and weaknesses. In particular, microscopy allows directly observing the organization of nanoparticles within the polymer matrix but is limited to a small portion of the studied material, thus requiring extensive sampling and statistical analyses to obtain representative and quantitative results. Furthermore, 2D images do not provide a direct view of the tri-dimensional organization in the matrix. X-ray micro-CT allows obtaining information on the 3D organization of the nanoparticles within a significant volumetric portion of the prepared nanocomposite, but with limited resolution and contrast. On the other hand, information on the filler content and dispersion within the polymer matrix on a large volume can be indirectly obtained from the evaluation of the rheological parameters of nanocomposites in the molten state [114]. In particular, the evolution of the complex viscosity with the frequency may provide insight on the percolation degree of nanoparticles in the polymer matrix [58, 114]. Furthermore, percolated nanocomposites usually exhibit a yield stress ($\sigma_0$) behavior, resulting from the hindrance of the polymer flow caused by the arrangement of nanoparticles in complex architectures, which may be correlated to the strength of the percolation network of GRM flakes within the polymer.

In the present study, the dispersion and organization of the nanoparticles in the polymer matrix was routinely characterized by FESEM and further investigated by rheology. Representative micrographs of cryo-fractured surfaces are reported in Figure 5. GRM nanoflakes are clearly embedded within the pCBT matrix, with generally good adhesion and reflecting the features of the GRM powders. Flat particles and apparently rigid flakes are clearly recognized in nanocomposites prepared with GRM-1, -3, -7, -8, -9, -10, in agreement with observation on the pristine powders. Particles size can hardly be evaluated in these conditions, but the dispersion in particles size appears to be very wide, likely due to the partial fragmentation of GRM flakes during compounding, particularly in the cases of larger initial size of GRM particles. Thinner and wavy flakes in the pristine GRM are less evident once embedded in the pCBT matrix (especially for GRM-5, -5T, -11) confirming excellent infiltration of pCBT within the galleries in accordion-like structure GRMs and the good particle-matrix adhesion. While contacts between overlapped particles can be observed in most of the reported micrographs, electron microscopy cannot directly provide insight on the 3D organization of particles and possible formation of a percolation network.

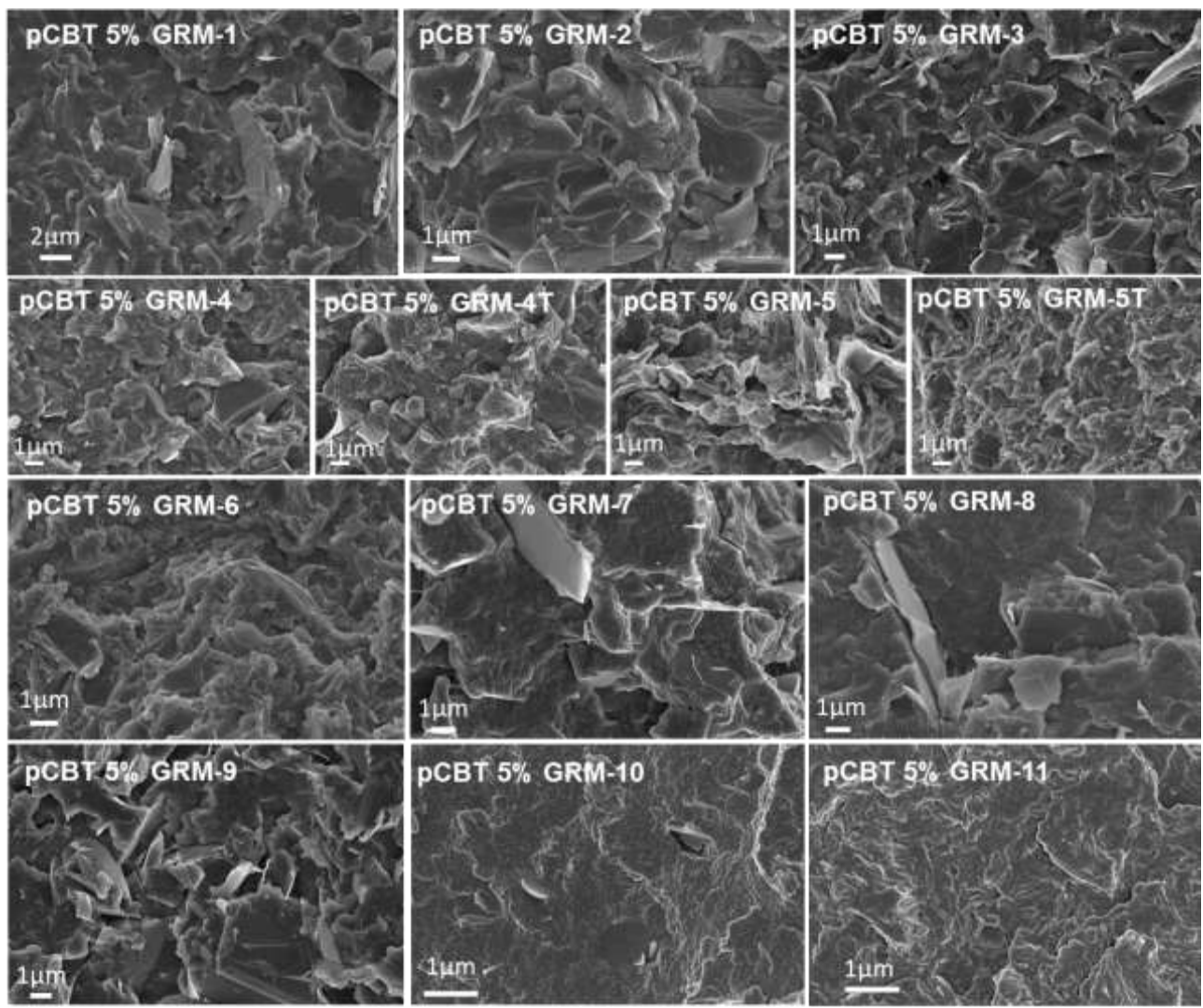

*Figure 5: Representative FESEM micrographs for the different pCBT/GRM nanocomposites*

In order to investigate the presence and properties of a percolation network of GRM flakes, rheological properties of the different nanocomposites were studied in details. Complex viscosity results *vs.* angular frequency for pCBT and its relevant nanocomposites are reported in Figure 7, showing dramatic differences between the different formulations. The nanocomposites containing GRM-7 and GRM-10 exhibit complex viscosity values lower than that of pristine pCBT in the whole frequency range explored. This reduction in the viscosity might be related to either a lubricant effect brought by relatively small and flat particles, as already proposed in literature [115], or to a lower molecular weight of pCBT in the presence of nanoflakes [56, 59]. Nanocomposite embedding GRM-8 also displayed a viscosity lower than pCBT and approx. constant in the mid and high frequency range, while a significant increase in viscosity is observed when decreasing frequency below 1 rad/s, thus suggesting the presence of a weak network of nanoparticles. In contrast, all other nanocomposites showed a strong dependence of the η* with the frequency, evidencing for a high percolation degree of nanoflakes within the polymer matrix (Figure 7). Dramatic increase in the complex viscosity at low shear rates are observable especially for the nanocomposites containing highly expanded and wavy flakes (GRM-4, -4T, -5, -5T, -6, -11), suggesting strong interaction between these flakes in the polymer matrix are maintained even at high shear rates. To get further insight on the quality of the percolation network, the yield stress for the different nanocomposites was calculated according to the modified Carreau model [116], obtaining an excellent fitting of the experimental data (Figure 6).

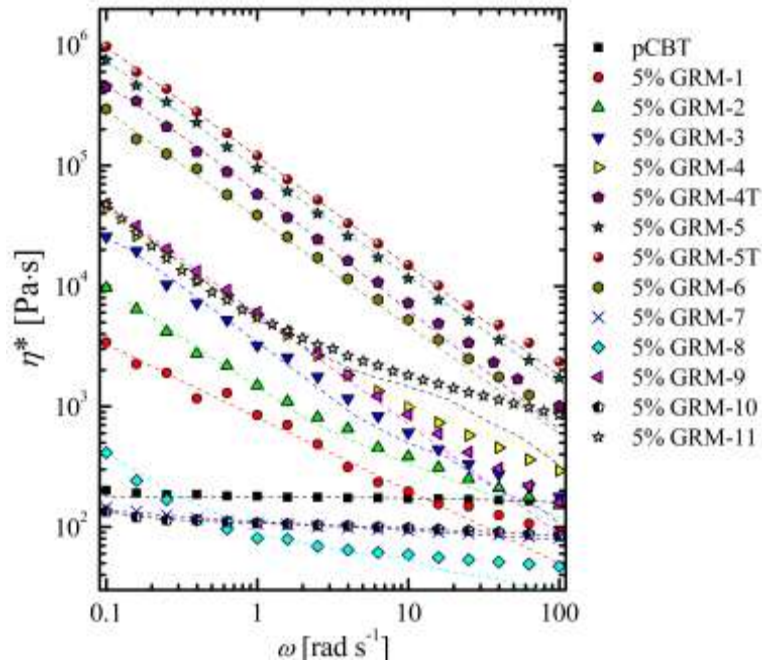

*Figure 6: Fitting of experimental complex viscosity data with modified Carreau model*

Nanocomposites showing highest viscosity (GRM-4T, -5, -5T, -6) also display high $\sigma_0$ values in the range 25 to 73 Pa, whereas limited (i.e. < 10 Pa) or negligible yield stresses are obtained from fitting of the viscosity curves of the pristine polymer and nanocomposites displaying low viscosity values (Table 2). A qualitative correlation between high yield stress values and the outstanding complex viscosity values at low shear is therefore apparent for nanocomposites based on highly expanded and wavy nanoplatelets. These observations demonstrate the presence of a rheologically strong percolation network, likely associated to the impregnation of polymer in the galleries of the GRM accordion-like structure, without the complete separation of individual nanoplates. This would results in domains of pre-organized nanoplates, able to interact between them and/or with dispersed individual nanoplates, strongly hindering the relaxation dynamics of the polymer macromolecules and the viscous flow of the matrix.

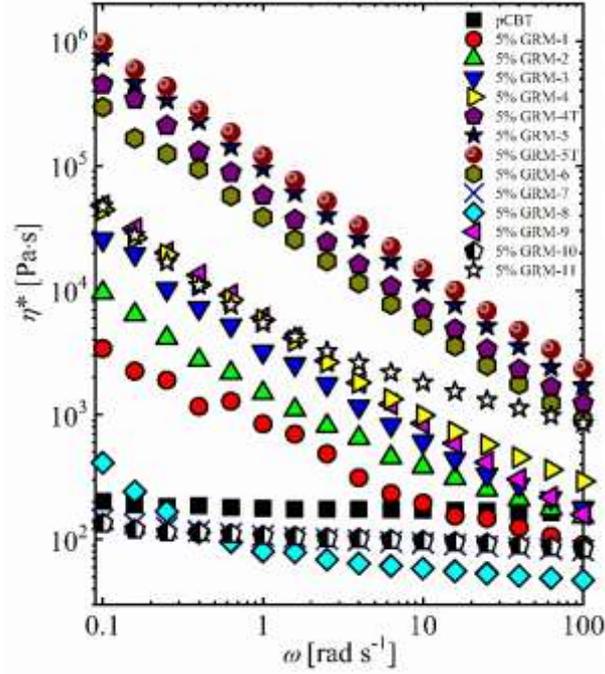

*Figure 7. Complex viscosity as function of the angular frequency for the different nanocomposites*

The presence of a percolating network of thermally conductive nanoparticles within an insulating polymer matrix is expected to enhance the heat transport properties of the relevant nanocomposites. Attempting to correlate the nanocomposites thermal conductivity with the quality of their GRM percolation network, λ values are reported *vs.* the complex viscosity at *ω* = 0.1 rad/s in Figure 8a. Nanocomposites exhibiting little or no increase in low shear rate viscosity (GRM-7, GRM-8 and GRM-10) led to a limited or no enhancement in the thermal conductivity, compared to the pristine polymer. This result is not surprising as the occurrence of a dense GRM percolating network is generally assumed to be needed for efficient heat transfer. On the other hand, the best heat conduction performance was obtained with GRM-5T, corresponding to the highest viscosity value. Beside these results, a general increasing trend for λ may be recognized as a function of viscosity, within a significant data scattering. Indeed, the several nanocomposites with thermal conductivity in the range of 0.4 to 0.6 W/mK (GRM-1, 4, 5, 7, 8, 9) correspond to extremely variable viscosity (@0.1 rad/s), ranging from $10^2$ to $10^5$ Pa·s. On the other hand, nanocomposites with $\eta^*_{@0.1\ rad/s}$ in the range of $10^5$ Pa·s (GRM-4T, 5, 5T, 6) correspond to rather variable λ values. The correlation between nanocomposites thermal conductivity and yield stress values (Figure 8b) provides similar trends, with a general increase in λ when increasing the $\sigma_0$ value, along with a wide scatter of experimental points. Despite the two rheological parameters ($\sigma_0$ and $\eta^*_{@0.1\ rad/s}$) identified in this work for the correlation with thermal conductivity appear to be qualitatively equivalent, it is worth noting that $\sigma_0$ is a fundamental parameter describing the rheological properties of the material, while the low shear rate viscosity value was arbitrarily taken at 0.1 rad/s.

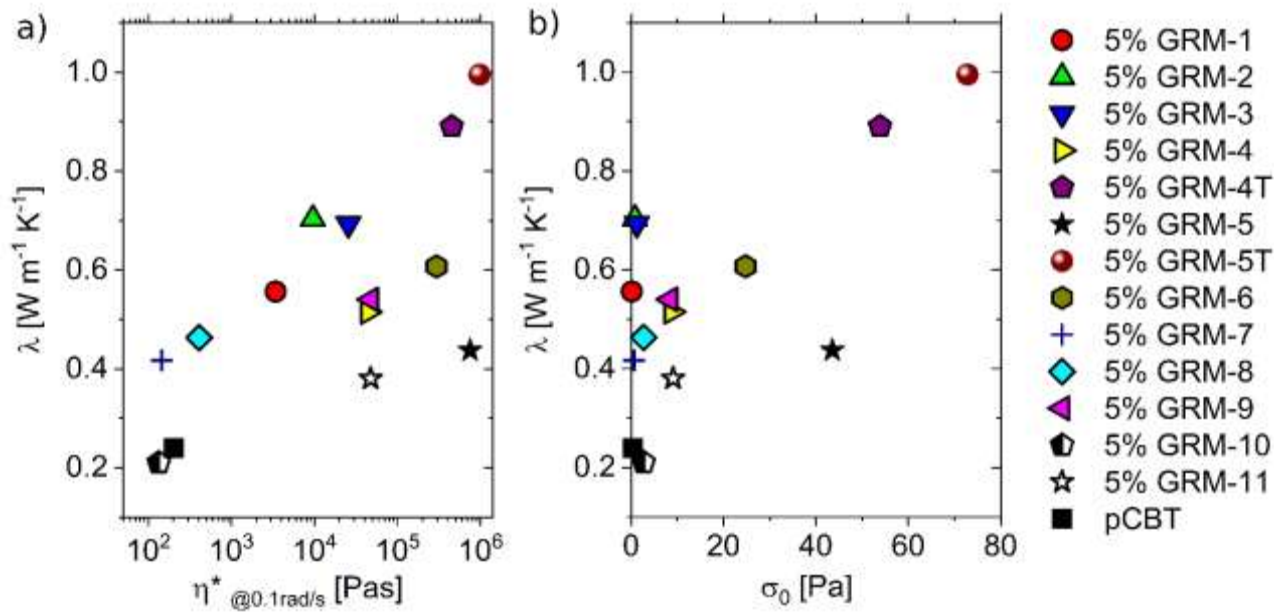

*Figure 8: Nanocomposite thermal conductivity vs. complex viscosity at 0.1 rad/s (a) and vs. yield stress (b)*

### 3.4. Combined effect of the different parameters

From the results above reported and discussed, none of the individual features (nanoparticles size and quality, degree of dispersion and percolation network strength) identified to affect the thermal conductivity in polymer nanocomposites appears to be sufficient as such to describe and quantify the thermal conductivity enhancement obtained by embedding graphene-related materials into a polymer. Despite some general trends can be observed, no univocal and quantitative correlation can be drawn between the thermal conductivity of the polymer nanocomposites and neither the defectiveness of the GRM, quantified as the $I_D/I_G$ ration in Raman scattering, nor the strength of the percolating GRM network within the nanocomposite, quantified by the low shear rate viscosity or the yield stress. These observations suggest the addressed parameters to be dependent with each other and the combination of such parameters should indeed be addressed. The thermal conductivity results for pCBT nanocomposites as function of the nanoflakes defectiveness and the complex viscosity at low shear rate are resumed in Figure 9, in which λ trends can be easily rationalized as a function of defectiveness and percolation network quality. When addressing GRM having similar defectiveness (e.g. GRM-5T vs GRM-8), the higher the viscosity of the nanocomposite melt, the greater the thermal conductivity obtained. On the other hand, at given viscosities, the lower the defectiveness the higher the thermal conductivity of the nanocomposites as, for instance, with GRM-5 vs GRM-5T.

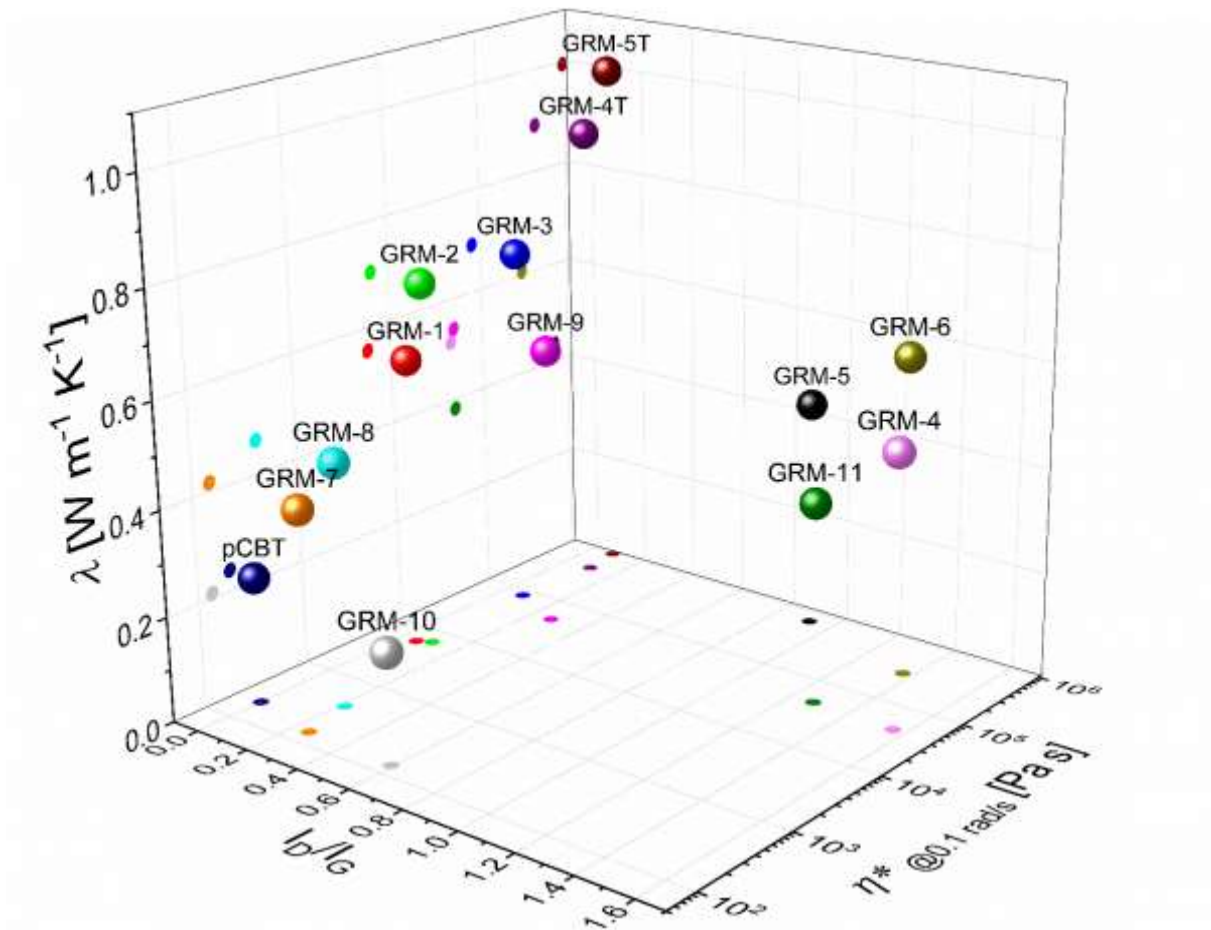

*Figure 9. Thermal conductivity of pCBT nanocomposites vs. GRM defectiveness and vs. melt viscosity (intrinsic, measured at 0.1 rad/s)*

## 4. Conclusions

13 different graphene-related materials with very variable features, including lateral size, thickness and defectiveness, were selected, characterized and exploited in the preparation of nanocomposites via melt mixing polymerization. Nanoplates were homogeneously dispersed within the polymer matrix, taking advantage of the infiltration of low viscosity cyclic oligomers followed by ring opening polymerization, leading to a set of nanocomposites with morphologies reflecting the structural features of as received GRM. The organization of such GRM nanoplates in a 3D network was investigated by rheological analyses, providing evidences for the presence of percolating networks within the prepared nanocomposites. However, dramatic differences were highlighted between different composite and related to the lateral size and morphology of the pristine GRM. Indeed, relatively small and flat nanoplates were found to produce limited impact on the viscosity, especially at high deformation frequencies, corresponding to weak percolation networks. On the other hand, highly expanded and wavy flakes were found to produce a strong intrinsic viscosity *vs.* frequency dependence, with viscosity increases up to 4 orders of magnitude at low shear rate as well as high yield stresses, evidencing for a high percolation degree of nanoflakes within the polymer matrix. Such network of GRM flakes is explained by the impregnation of polymer in the galleries within accordion-like structure of highly expanded GRM grades, without the complete separation of individual nanoplates upon melt mixing and polymerization. Beside affecting the viscous flow of the matrix, the density and strength of the network was confirmed to affect the thermal conductivity of the nanocomposites. However, the comparison of different nanocomposites, displaying similar rheological properties, demonstrated that the quality of the network is not sufficient to univocally describe and quantify the thermal conductivity enhancement. Similarly, the defectiveness of GRM nanoplates was found to affect the thermal conductivity without being sufficient as such to determine the thermal conductivity performance in nanocomposites. The combination between quality of the conductive particles and quality of the network was indeed demonstrated to determine the thermal conductivity in polymer nanocomposites. This result highlights the need of a careful selection of low-defectiveness conductive particles and processing design, to retain extensive contact area between nanoplates and minimize thermal resistances, when aiming at highly efficient thermally conductive materials.

**Acknowledgment**

This work has received funding from the European Research Council (ERC) under the European Union's Horizon 2020 research and innovation programme grant agreement 639495 — INTHERM — ERC-2014-STG and previously from the European Union Seventh Framework Programme, under grant agreement n° 604391 "Graphene Flagship".

Julio Gomez at Avanzare Innovación Tecnólogica- E and Francescco Bertocchi at Nanesa-I are gratefully acknowledged for providing GRM materials. Mauro Raimondo at Politecnico di Torino is also acknowledged for FESEM observations.

**Competing Interest**

The authors declare no competing interest.

The comparison of results obtained with different grades of GRM in this paper is not aimed at assessing which production process or which commercial product is best, but exclusively to investigate and define scientific correlations between GRM structural features and nanocomposites thermal conductivity.

**Author contributions**

A.F conceived this research work and the experiments within, interpreted the experimental results and led the research activities. S.C. carried out most of the nanocomposites preparation and characterization. D.B. contributed in nanocomposites preparation, characterization and data elaboration. M.E. carried out Raman characterization of GRM. R.A. contributed in the elaboration and interpretation of rheological results. Manuscript was mainly written by A.F and S.C.